\documentclass[11pt, a4paper]{article}
\usepackage{hyperref}
\usepackage{graphicx}
\usepackage[hscale=0.75,vscale=0.8]{geometry}
\usepackage{amssymb}
\usepackage[small,bf]{caption}

\begin{document}

\title{Entanglement entropy and algebraic holography}

\author{Bernard S. Kay$^*$ \medskip \\  {\small \emph{Department of Mathematics, University of York, York YO10 5DD, UK}} \smallskip \\ 
\small{$^*${\tt bernard.kay@york.ac.uk}}}

\date{}

\maketitle

\begin{abstract}
In 2006, Ryu and Takayanagi (\emph{RT}) pointed out that (with a suitable cutoff) the entanglement entropy between two complementary regions of an equal-time surface of a $d$+1-dimensional conformal field theory on the conformal boundary of AdS$_{d+2}$ is, when the AdS radius is appropriately related to the parameters of the CFT, equal to $1/4G$ times the area of the $d$-dimensional minimal surface in the AdS bulk which has the junction of those complementary regions as its boundary, where $G$ is the bulk Newton constant.  (More precisely, RT showed this for $d=1$ and adduced evidence that it also holds in many examples in $d>1$.)   We point out here that the RT-equality implies that, in the quantum theory on the bulk AdS background which is related to the boundary CFT according to Rehren's 1999 algebraic holography theorem, the entanglement entropy between two complementary bulk Rehren wedges is equal to one $1/4G$ times the (suitably cut off) area of their shared ridge.  (This follows because of the geometrical fact that, for complementary ball-shaped regions, the RT minimal surface is precisely the shared ridge of the complementary bulk Rehren wedges which correspond, under Rehren's bulk-wedge to boundary double-cone bijection, to the complementary boundary double-cones whose bases are the RT complementary balls.)   This is consistent with the Bianchi-Meyers conjecture -- that, in a theory of quantum gravity, the entanglement entropy, $S$ between the degrees of freedom of a given region with those of its complement takes the form $S = A/4G$ (plus lower order terms) -- but only if the phrase `degrees of freedom' is replaced by `matter degrees of freedom'.   It also supports related previous arguments of the author -- consistent with the author's `matter-gravity entanglement hypothesis' --  that the AdS/CFT correspondence is actually only a bijection between just the matter (i.e.\ non-gravity) sector operators of the bulk and the boundary CFT operators.  
\end{abstract}

\maketitle

The main purpose of this letter is to point out a hitherto apparently unnoticed interconnection between two developments, both of which have to do with the AdS/CFT correspondence \cite{maldacena1997large, gubser1998gauge, witten1998anti} (for AdS of any dimension $d$+2):  the \textit{algebraic holography theorem} \cite{rehren2000algebraic, rehren2000local, rehren2004qft} of Rehren and the \emph{RT-equality}, i.e.\ the equality \cite{ryu2006holographic, ryu2006aspects} due to Ryu and Takayanagi, which relates the entanglement entropy of complementary spatial regions of a conformal field theory (\textit{CFT}) defined on the AdS conformal boundary to the area of a certain related minimal surface in the AdS bulk.

Rehren's Algebraic Holography theorem is based on a naturally defined geometric bijection between suitably defined spacelike wedges\footnote{We shall call the wedges involved in the Rehren bijection `Rehren wedges' here to distinguish them from the more general wedges considered e.g.\ by \cite{hubeny2007covariant, hubeny2012causal, hubeny2013global} which include wedges dual to boundary double cones whose bases are more general shapes than balls and we shall sometimes refer to a double cone and the Rehren wedge which is its image under the Rehren bijection as being \textit{dual} to one another.} in the AdS bulk and double cones on the AdS conformal boundary.  As explained in \cite{rehren2000algebraic, rehren2000local, rehren2004qft}, this induces a bijection between the net of local quantum algebras for a CFT defined on the conformal boundary and the net of local algebras for an AdS-invariant quantum theory (which we shall call the \textit{Rehren dual} of the CFT) on the AdS bulk which satisfies a suitable notion of commutativity at spacelike separation.  

Leaving aside, temporarily, the need for a suitable cutoff, the basic RT-equality -- which, in \cite{ryu2006holographic, ryu2006aspects}, was demonstrated  for $d$=1 and conjectured, on the basis of semi-qualitative evidence in several examples, for $d>1$ --  is between the entanglement entropy of two complementary connected open subsets of an equal-time surface for a CFT on the AdS conformal boundary on the one hand and $1/4G$ times the area of the minimal $d$-dimensional surface in the AdS bulk which reaches the conformal boundary at the junction of the two complementary open subsets\footnote{Here, we say a pair of open subsets is complementary if one is the complement of the closure of the other and we then call their common boundary their \textit{junction}}, as we next explain in the case of interest here where the complementary subsets are complementary balls.  $G$ here denotes the appropriate bulk Newton constant, and the important proviso needs to be made that the RT equality only holds when $G$ is related to the AdS radius, $R$, and to the parameters of the boundary CFT as in the AdS/CFT correspondence.   In $d=1$, the appropriate relation \cite{ryu2006holographic, ryu2006aspects} is the Brown-Henneaux  \cite{brown1986central} relation,
\begin{equation}
\label{BrownHenneaux}
G={3R\over 2c},
\end{equation}
where $c$ is the CFT's central charge.    In $d>1$, there is no such general formula; there is a different relation for each AdS/CFT model which depends on the details of the model.  For the sake of brevity, we will refer collectively to the $d=1$ and $d>1$ relations as the \textit{BH relation}.  We also recall that $R$ is related to the cosmological constant, $\Lambda$ by
\begin{equation}
\label{R-Lambda}
\Lambda=-{d(d+1)\over 2R^2}.
\end{equation}
To make all this precise and illustrate it, we focus first on the $d$=1 case: AdS$_3$ can be considered to be the open solid cylinder of radius $\pi/2$ of Figure 1 (cf. Figure 1 in \cite{rehren2000local} and Figure 1 in \cite{ryu2006holographic}).   

\begin{figure}
   \centering
    \includegraphics*[scale = 1.5]{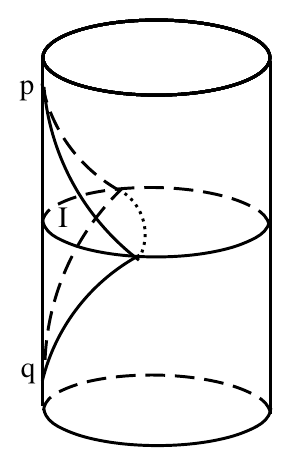}
   \caption{1+2-dimensional AdS showing an-equal time circle, a typical interval, $I$, of that circle, the boundary double cone which has $I$ as its base and the Rehren wedge dual to that double cone.  The dotted arc is the `ridge' of the Rehren wedge.}  
 \end{figure}

Choosing global coordinates, $(t,r, \theta)$, $t\in (-\infty, \infty)$, $r\in [0, \pi/2)$, 
$\theta \in [0, 2\pi)$\footnote{\label{coordinates}The global coordinates of our main text are related to the global coordinates,  $(t,\rho, \theta)$ referred to in \cite{ryu2006holographic, ryu2006aspects}, which uses the opposite signature to us and in which the metric takes the form $ds^2=R^2(-\cosh^2\!\!\rho\, dt^2 + d\rho^2 + \sinh^2\!\!\rho\, d\theta^2)$,  by $\sec r = \cosh\rho$.} in which the metric takes the form
\[
ds^2=R^2\sec^2\!r(dt^2-dr^2-\sin^2\!r\, d\theta^2),
\]
where $R$ is the AdS radius,  the AdS conformal boundary is then identified with the boundary of the cylinder at $r=\pi/2$.   For $d>1$ one replaces $d\theta^2$ by the usual metric on the $d$-sphere.  

We remark that this cylinder (and each of its counterparts for $d>1$) represents unwrapped AdS -- i.e.\ the covering space of true AdS -- whereas Rehren's theorem applies to the quotient of wrapped -- i.e.\ true --  AdS  by the equivalence relation which identifies antipodal points.   As pointed out in \cite{rehren2000algebraic}, this has, as its conformal boundary, the conformal compactification of Minkowski space --  which is the natural domain of definition for conformal field theories.   For many purposes, however (and as indeed is also said in \cite{rehren2000local}) it is not too misleading, and can help to simplify several explanations, if one refers, as we will here, to the unquotiented covering space.  (Ordinary, uncompactified) Minkowski space is then to be identified with the intersection with the AdS conformal boundary of the domain of a choice of Poincar\'e chart (whose definition we shall recall below).  See e.g.\ Endnote [8] in \cite{kay2008pre} for more discussion.

In the conformal boundary of AdS, a double cone consists of the intersection of the backward light cone of a point $p$ with the future light cone of a point $q$ to its past.  A typical such cone in the conformal boundary of AdS$_3$ (see Figure 1) consists of the domain of dependence, in the conformal boundary, of an interval $I$, say (if the interval is entirely contained within a single coordinate chart) $\theta \in (\theta_1, \theta_2)$, of the circle located at the intersection of a constant-$t$ surface with the conformal boundary $r=\pi/2$ -- the general double cone being the result of acting on such a double-cone with a conformal transformation.   For $d>1$ the circle is replaced by a $d$-sphere and a typical double-cone is the domain of dependence, in the conformal boundary, of a \textit{ball} i.e.\ of a subset of the $d$-sphere consisting of points whose geodesic distance in the $d$-sphere from one of its points is less than some value no bigger than the $d$-sphere semi-circumference.  We shall then call the ball (or, if $d=1$, the interval) the \textit{base} of the double-cone.  The \textit{Rehren wedge} which is dual to such a double-cone is \cite{rehren2000local} defined to be the intersection, now in the AdS bulk, of the interior of the past light cone of $p$ with interior of the future light cone of $q$.  We shall say that $p$ is the future \textit{vertex} and $q$ the \textit{past vertex} of that Rehren wedge.

For a given CFT on the conformal boundary of AdS$_3$ in its natural ground state and a given interval of our $t=0$, $r=\pi/2$ circle, the RT equality is the statement that the entanglement entropy of an interval and its complementary interval (i.e.\ the interior of the set-theoretic complement of that interval in the circle) is equal to $1/4G$ times the length of the bulk geodesic which reaches the conformal boundary at (what we shall call here) the \textit{junction} of the complementary intervals -- i.e.\ at the set consisting of their two shared endpoints.  This generalizes to $d>1$: The RT equality equates  the entanglement entropy of complementary balls in the $t=0$ $d$-sphere on the AdS$_{d+2}$ conformal boundary with $1/4G$ times the area of the bulk minimal surface which reaches the conformal boundary at the \textit{junction} of these complementary balls  -- i.e.\ at the $d$-1 sphere which is their common boundary.

The connection with Rehren holography springs from \textit{the simple geometrical fact that the above RT minimal surface is precisely the same as the shared} ridge \textit{of the Rehren wedges which, in Rehren's algebraic holography, are dual to the boundary double-cones whose bases are the complementary balls}.\footnote{Note that it is important for this result that the wedges are Rehren wedges and not the more general wedges referred to in Footnote 1.   See the papers \cite{hubeny2007covariant, hubeny2012causal, hubeny2013global} for an explanation of why the result does not generalize to more general wedges.} Here, the \textit{ridge} of a Rehren wedge\footnote{We borrow the term `ridge' from Rehren's paper \cite{rehren2000algebraic}.  In the case $d=0$ (which is not of relevance here) , where the conformal boundary is one-dimensional and the ridge a single point, Rehren calls it the `apex'.} is defined to be the intersection of the boundary of the Rehren wedge with the interior of the bulk.  In other words, it is the intersection, in the AdS bulk, of the past light cone of the Rehren wedge's future vertex with the future light cone of its past vertex.   This is illustrated in Figure 1 for the case $d=1$.

To prove that the ridge of a Rehren wedge is a minimal surface, it is convenient to choose a Poincar\'e chart whose domain contains that Rehren wedge (and the boundary of whose domain contains the boundary double-cone which is dual to that Rehren wedge).  Here we recall that the domain of a Poincar\'e chart is a maximal Rehren wedge, dual to (i.e.\ reaching the conformal boundary at) a boundary double-cone whose base is a maximal ball -- consisting of our $t=0$ $d$-sphere with a single point removed.    Poincar\'e coordinates, $(t, x_1 \dots x_d; z)$ map this domain to the $d$+2-dimensional half-space $\mathbb{R}^{d+1}\times (0,\infty)$, and, in these coordinates, the bulk metric takes the form 
\begin{equation}
\label{Poincare}
ds_{\mathrm{Poincare}}^2=\frac{R^2}{z^2}(dt^2 - dx_1^2 - \dots - dx_d^2- dz^2)
\end{equation}
which is conformal to the Minkowski metric
\begin{equation}
\label{Minkowski}
ds^2_{\mathrm{Minkowski}}=dt^2 - dx_1^2 - \dots - dx_d^2 - dz^2.
\end{equation}
 -- the conformal boundary being located at $z=0$.    Suppose the future/past vertices of the given Rehren wedge have Poincar\'e coordinates $(t,x_1, \dots, x_d; z) = (\pm T, 0,\dots, 0; 0)$.   Then, since the light cone is unaffected by the conformal factor $\frac{R^2}{z^2}$ and therefore is the same as the light cone in the Minkowski metric (\ref{Minkowski}),  the intersection of the two light cones within our half-space will obviously  be the $d$-hemisphere (in $d=1$, semicircle) $\{ (0,x_1, \dots x_d; z) \in R^{d+1} \times [0,\infty)| x_1^2 +\dots x_d^2 + z^2 = T^2\}$ in the $t=0$ plane which, moreover, obviously meets the boundary at right-angles.   But the $t=0$ plane with the metric obtained by deleting $t$ in (\ref{Poincare}) is a well-known representation of $d$+1-dimensional hyperbolic space (with radius $R$) -- the generalization to general $d$ of the usual upper half plane representation of 2-dimensional hyperbolic space in the case of $d=1$ -- in which it is well known that the minimal surfaces are precisely the hemispheres which meet its boundary at right-angles.  (In the case $d=1$, the geodesics are precisely the semi-circles which meet its boundary at right-angles.)  

The gist of the main observation which we wish to make in this letter is that the following sequence of identities holds:   First, the quantum algebra of a pair of complementary balls on the conformal boundary at a fixed time must --  if we assume the boundary CFT algebra satisfies the time-slice condition -- be the same as the quantum algebra of the pair of complementary boundary double-cones which have these balls as their bases.  This, in its turn, must, by Rehren's algebraic holography theorem, be isomorphic to the Rehren quantum algebra of the pair of complementary Rehren wedges which are dual to these double-cones.  (And the restriction of the ground state to the pair of complementary boundary double-cones will map, under the induced map on states, to the restriction of the bulk ground state to the complementary Rehren wedges.)  As an immediate consequence of this isomorphism, it must follow, in particular, that the entanglement entropy between the complementary balls  in the $t=0$ $d$-sphere in the boundary CFT (in its ground state) must be the same as the entanglement entropy (in the bulk Rehren quantum theory in its bulk ground state) between the complementary Rehren wedges. Furthermore, in view of the RT equality, and, in the presence of the BH relation, the entanglement entropy of the complementary Rehren wedges must equal $1/4G$ times the area of the RT-minimal surface which, by the above simple geometrical fact, is the same thing as $1/4G$ times the area of their shared ridge.    

Of course, none of this means much in the absence of a cutoff:   One expects both of the  (uncut-off) entanglement entropies mentioned above to be infinite and also the area of the minimal surface (/shared ridge) will be infinite.   However, these infinities are, of course, also present in the original RT work described in \cite{ryu2006holographic, ryu2006aspects} and those papers show how (following \cite{susskind1998holographic}) to introduce a cutoff -- which is at the same time an ultra-violet cutoff for the boundary CFT and a large-volume cutoff for the bulk theory -- which renders the RT equality meaningful.  Basically (see \cite{ryu2006holographic, ryu2006aspects} for details) one replaces the boundary cylinder located at $r=\pi/2$ by a cylinder drawn inside the bulk at a slightly smaller $r$-coordinate, say $r_0$, and replaces the $n$-point functions of the CFT on the boundary cylinder by the $n$-point functions on that slightly smaller cylinder.  Also one takes the cut-off area of the RT-minimal surface (for $d=1$, length of the RT-geodesic) to be the (finite) area of its portion inside the same slightly smaller cylinder. 
 
The true RT equality is then between the (now finite) entanglement entropy for our complementary balls in the cutoff CFT\footnote{For $d=1$ (see e.g.\ Equations (1.3), (2.3) and (2.5) in \cite{ryu2006holographic} and Footnote \ref{coordinates} here) the 
cut-off entropy between complementary intervals of length $\theta$ and $2\pi-\theta$ may then be interpreted as the entanglement entropy between complementary intervals of length $l$ and $L-l$ for a 1-dimensional quantum many-body system of length $L$ with periodic boundary conditions, described at criticality by our CFT, with lattice spacing (or ultra-violet cutoff) $a$ when $a$ and $L$ are related to $r_0$ by $\pi/2-r_0 \sim 2a/L$ and $\theta$ is identified with $2\pi l/L$.} on the conformal boundary and $1/4G$ times the latter finite area.  

We propose to apply this same cutoff to Rehren's algebraic holography map:   To illustrate how this might be done, suppose, for example, that the bulk Rehren theory is a Wightman theory involving a scalar field $\phi(t,r,\theta)$ where, for $d>1$, $\theta$ denotes a point on the $d$-sphere.   Then, the usual boundary algebra on the conformal boundary at $r=\pi/2$ will be obtained \cite{kay2008pre, bertola2000general} by defining boundary fields, $\phi_b$, with
\[
\phi_b(t,\theta)=\lim_{r\rightarrow \pi/2} (\cos r)^{-\Delta} \phi(t,r,\theta)
\]
where $\Delta$ is a suitable constant (which becomes the anomalous dimension of the 
resulting boundary CFT).   For our cut-off boundary algebra, we propose, instead, to define the \textit{cut-off boundary fields}, $\phi_c$, on, say, now the $t=0$ boundary sphere, by
\[
\phi_c(\theta)= (\cos r_0)^{-\Delta}\phi(t=0, r_0,\theta)
\]
and to define the cut-off algebra on the boundary sphere at $t=0$ and $r=\pi/2$ to be the field algebra generated by these.  Note that it is important that we regard this algebra as located on the boundary sphere at $r=\pi/2$ and \textit{not} at $r=r_0$.  Assuming again that  the time-slice condition holds for the boundary CFT, we declare the full boundary cut-off algebra to be identical with this latter $t=0$, $r=\pi/2$ cut-off algebra.  Further, for any ball-shaped region in the boundary sphere at $t=0$ and $r=\pi/2$, we declare the cut-off algebra for the boundary double-cone which has that ball-shaped region as its base to be identical with the subalgebra of the $t=0$ cut-off algebra generated by $\phi_c(\theta)$ for $\theta$ located in that ball-shaped region.   Finally we declare the cut-off Rehren algebra for the bulk Rehren wedge dual to such a double cone to equal the cut-off algebra of its dual boundary double-cone.   But we regard that bulk wedge as extending only as far as $r=r_0$ so that, in particular, the area of its ridge will be finite.

We note that the cutoff breaks conformal invariance on the conformal boundary and (therefore) also breaks invariance under the AdS group in the bulk.  However it respects time-translation invariance both in boundary and bulk.   

However,  we will still have a net isomorphism between the net of double-cone (now cut-off) boundary algebras and the net of (now cut-off) algebras for our (cut-off) bulk Rehren wedges (albeit the boundary theory will not be exactly conformally invariant and the bulk theory will not be exactly invariant under the AdS isometry group).  

Assuming this cutoff to be in place we can now make precise our main observation, the gist of which we outlined above:

\medskip

\noindent
\textbf{Observation O.} \textit{With our cutoff, the entanglement entropy of a pair of complementary balls on the conformal boundary at a fixed time will be finite and, assuming the time-slice condition holds for the boundary theory, equal to the entanglement entropy of the pair of boundary double cones which have these balls as their bases.  Moreover,  this will  equal the entanglement entropy of the complementary Rehren wedges in the bulk which correspond to these complementary balls under the Rehren bijection.} 

\medskip

Furthermore, by the RT equality and the above simple geometrical fact, we may conclude that the bulk Rehren theory must have the property:

\medskip

\noindent
\textbf{Property P.} \textit{The entanglement entropy of the cut-off algebras for a pair of complementary Rehren wedges is (in the presence of the BH relation) equal to $1/4G$ times the cut-off area of their shared ridge.}  

\medskip

We remark immediately that, were we to have an independent explanation for Property P (for which see the end of Point (5) below) then, since, by the above simple geometrical fact, the shared ridge is the same thing as the RT minimal surface for our pair of complementary ball-shaped regions, this would provide an explanation of the RT equality (in the case of ball-shaped complementary regions).  

We next wish to fill in some details related to this observation and to make some further remarks and draw some further conclusions about its significance:

\medskip

(1) For a given Rehren wedge there will be a one-parameter subgroup of the AdS isometries which maps the wedge into itself and the restriction of the AdS ground state to that wedge will be a KMS state -- i.e.\ thermal equilibrium state\footnote{Note that what one considers to be the temperature of this thermal state will of course depend on an arbitrary choice of scale and is unimportant.} -- with respect to it.   In fact \cite{rehren2000algebraic} one may identify the one-parameter subgroup with the modular group -- in the sense of Tomita-Takesaki -- of the restriction of the ground state to the wedge's von Neumann algebra, while the modular involution will map the von-Neumann algebra for the Rehren wedge to the von-Neumann algebra for the complementary Rehren wedge and the KMS property is an immediate corollary to the former statement.   The entanglement entropy between the wedge and its complementary wedge is of course the same thing as the entropy of that thermal state.   In the absence of a cutoff it will be infinite.

\medskip    

(2) Property P strongly suggests that, \textit{in the uncut-off bulk Rehren theory},  the entanglement entropy \textit{per unit area of the shared ridge} of a pair of complementary bulk Rehren wedges is (in the presence of the BH relation) equal to $1/4G$.

\medskip

(3) It is not a surprise that this entropy should be proportional to the area of the shared ridge.  After all, computations  going back to Bombelli et al.\ \cite{bombelli1986quantum} (cf. also e.g.\ \cite{tHooft1985quantum, mukohyama1988black} for the analogous situation in the Kruskal spacetime) indicate that, say, for an ordinary free field in Minkowski space in its usual vacuum state, the entropy per unit area of the shared ridge for a pair of Rindler wedges, is, with a suitable short-distance regularization\footnote{not to be confused with our cutoff}, a finite constant. However it may seem to be a surprise that our entanglement entropy per unit area is finite since we have not imposed any short-distance cutoff on our bulk theory.   By contrast, in the analogue systems just mentioned, on removal of the short-distance regularization, it becomes infinite. 

\medskip

(4) The fact that the entanglement entropy of a pair of complementary balls on the AdS conformal boundary is equal to the entanglement entropy of a pair of regions (namely the pair of complementary bulk Rehren wedges discussed above) for a quantum theory on a fixed AdS background appears to be in conflict with the conventional view that the AdS/CFT correspondence is an equivalence between a CFT on the conformal boundary of AdS with a full theory of quantum gravity in the bulk.   For, surely, were this conventional view to be valid, one would expect it, instead, to be equivalent to an entanglement entropy between two subsystems of the full bulk quantum gravity theory which are, in some suitable sense, complementary.   This particular apparent conflict in fact serves to illustrate, and put into sharp focus, the more general apparent conflict, pointed out in 2001 by Arnsdorf and Smolin \cite{Arnsdorf2001Maldacena}, between the conventional view about AdS/CFT and the fact that Rehren's algebraic holography establishes an equivalence between the boundary CFT and a bulk quantum theory defined on the fixed AdS background.   For, as Arnsdorf and Smolin point out, this would immediately seem to imply that full quantum gravity in the AdS bulk is equivalent to a quantum field theory on a fixed AdS background.  It is difficult to see how this could be the case because the latter satisfies a version of commutativity at spacelike separation, whereas, one would not expect a full quantum gravity theory to satisfy such a property (whatever one might consider to be the background metric).  Arnsdorf and Smolin themselves considered a number of possible resolutions of this puzzle and amongst the possibilities they considered was that the AdS/CFT correspondence may be not a bijection but rather a one way map from bulk to boundary.  In \cite{kay2014brick, kay2015instability, kay2015nonexistence} the present author gave a number of arguments (based on consideration of the AdS/CFT correspondence when the bulk may contain a black hole -- see Point (6) below)  which support this view and in fact hypothesised that what is actually the case is that the boundary CFT is equivalent to a subtheory of the bulk quantum gravity theory, namely its `matter' sector -- it being plausible that this, in its turn, will be well approximated by a certain quantum theory involving matter on a fixed AdS (or, when a black hole is present, Schwarzschild-AdS) background.  The success of the Rehren theory in being able to account for boundary entanglement entropy as a consequence of Property P would seem to lend further support to this hypothesis and suggests that (in the case of AdS) the Rehren dual to the boundary CFT \textit{is} that certain quantum theory.

\medskip

(5) Assuming the validity of this latter suggestion,  we can immediately conclude, in view of Point (2) above, that the \textit{the entanglement entropy per unit area of shared ridge of the matter degrees of freedom of full quantum gravity for a pair of complementary bulk Rehren wedges is (in the presence of the BH relation) equal to $1/4G$}.

\medskip

This statement resembles a special case of the general conjecture due to Bianchi and Myers that (altering the quoted text only for consistency of notation)

\medskip

``\textit{In a theory of quantum gravity, for any sufficiently large region in a smooth background spacetime, the entanglement entropy between the degrees of freedom describing the given region with those describing its complement is finite and to leading order, takes the form $S=A/4G$ (plus lower order terms).}''

\medskip

\noindent
but suggests that this Bianchi-Myers conjecture ought to be modified by replacing the phrase ``the degrees of freedom'' by ``the \textit{matter} degrees of freedom''.   As far as we aware, all the evidence adduced in \cite{bianchi2014architecture} for the correctness of the conjecture may equally be taken to be evidence for its correctness when thus modified.   Furthermore, if we assume the validity of our suggestion at the end of Point (4), then,  in view of our remark in the previous sentence, the evidence adduced in \cite{bianchi2014architecture} provides us with an independent explanation for Property P.  (See the remark immediately following the statement of Property P above.)

\medskip

(6) As we indicated in Point (4) above, we have argued elsewhere (in \cite{kay2014brick, kay2015instability, kay2015nonexistence}) that the boundary CFT is equivalent just to the matter subsector of quantum gravity in the AdS bulk.  In this point, we wish to briefly recall our argument for this, which is based on consideration (see the next paragraph) of the AdS/CFT correspondence when the bulk contains a black hole.   Let us take as our starting point the statement (familiar from mainstream work on AdS/CFT) that, \textit{in AdS correspondences when the bulk contains a black hole, the states of the boundary CFT and of quantum gravity in the bulk are both thermal states and the entropy of the thermal state of the boundary CFT is equal to the entropy of the thermal state of quantum gravity in the bulk}.   We do not dispute the correctness of this statement, and we do not dispute what is meant by a thermal state of the boundary CFT or what is meant by its entropy.    But we disagree with the traditional view as to what is meant by a thermal state of quantum gravity in the bulk and about what is meant by its entropy.  On the traditional view, the total thermal state of quantum gravity in the bulk is a mixed state and what one means by its entropy is the von Neumann entropy of this total mixed state.    On our view, which is in accordance with our wider \textit{matter-gravity entanglement hypothesis} (see e.g.\ \cite{kay2015entropy} and references therein) a total thermal state of quantum gravity in the bulk is to be understood mathematically as a pure state, entangled between matter and gravity in just such a way that the reduced states of matter and of gravity separately are approximately thermal states and its entropy is the matter-gravity entanglement entropy of that total pure state -- which, by the way, is the same thing as the von Neumann entropy of the reduced state of matter (and also, by the way, the same thing as the von Neumann entropy of the reduced state of gravity). 

To provide our reasons for taking this view -- additional to reasons we have given \cite{kay2015entropy} for our matter-gravity entanglement hypothesis in general -- we begin by considering how algebraic holography would generalize to the maximally extended Schwarzschild-AdS spacetime and considering the extent to which this can have a counterpart in full AdS/CFT.  As is well known (see the discussion and references in \cite{kay2014brick} for details)  the right and left Schwarzschild wedges of the classical Schwarzschild-AdS spacetime each have their own separate conformal boundary cylinders and one expects there to be a natural Hartle-Hawking-Israel like state on the relevant bulk Rehren theory which is dual in a sense similar to that of Rehren to a total state of a CFT on the union of these two boundary cylinders which is pure, but entangled between the two cylinders in just such a way that it is separately thermal on each. (I.e.\ is a `double KMS state' in the sense of \cite{kay1985purification} [see also \cite{kay1985uniqueness}] .)  When the boundary CFT is chosen to be one for which there exists an AdS/CFT correspondence, and when the relevant BH relation holds, we expect (cf. the italicized sentence in the previous paragraph -- and see e.g. the summary of the standard literature on this topic in Endnote (i) in \cite{kay2014brick}) that the total entropy of the thermal state of the  boundary CFT on either of the cylinders will be given by the Hawking formula: $1/4G$ \textit{times the area of the Schwarzschild-AdS event horizon}.   (It is noteworthy that an alternative argument for this latter result --  as explained in Section IV of \cite{ryu2006holographic} --  is to notice that it arises as a limiting case of  the RT-equality.)  

By the arguments in \cite{kay2014brick}, we also expect that the entropy of the thermal state on (say) the right wedge of the bulk Rehren theory will also equal $1/4G$ \textit{times the area of the Schwarzschild-AdS event horizon} and hence, assuming the correctness of the conclusion at the end of Point (4) above, we expect that, in full AdS/CFT, it must be the entropy of the reduced state of the matter in the right wedge which will equal  $1/4G$ \textit{times the area of the Schwarzschild-AdS event horizon}.   In contrast, in \cite{maldacena2003eternal}, Maldacena proposes that this is the entropy of the reduced state of matter-gravity for the right wedge:  In fact, he assumes that in the AdS/CFT correspondence, when there is a black hole in the bulk, the CFT on the union of right and left boundary cylinders will be equivalent to a full theory of quantum gravity on the right and left Schwarzschild wedges respectively with the thermal states of the CFTs being the boundary limits of thermal states of full quantum gravity in each wedge which are, in their turn, reduced states of a pure total state of quantum gravity on the full maximally extended Schwarzschild-AdS.   

The picture arrived at in our own papers (in \cite{kay2014brick, kay2015instability, kay2015nonexistence}) is very different:  Not only do we argue to equate $1/4G$ \textit{times the area of the Schwarzschild-AdS event horizon} with the entropy of just the matter on the right wedge, but we argue in these papers that (unlike the situation for quantum field theory on a fixed spacetime background) if one attempts to define quantum gravity on full maximally extended Schwarzschild-AdS, the horizon will be unstable, and (in any relevant AdS/CFT correspondence) the closest thing possible to a classical description of a spacetime which is dual to a thermal state of the boundary CFT on a (now single) boundary cylinder (say the `right' cylinder) will resemble just the (say) right wedge spacetime of Schwarzschild-AdS but with the classical description breaking down near the horizon.  Furthermore the actual total state of quantum gravity which will have this approximate classical spacetime description will be a pure state, entangled between matter and gravity in just such a way that the reduced state of the matter (and also of the gravity) will be a thermal state at the Hawking temperature -- it being the matter sector (with its thermal state) only which is equivalent to the boundary CFT on the right cylinder (with \textit{its} thermal state).    The assumption here is that the (thermal) reduced state of the matter obtained by taking the partial trace over gravity of this total pure state of quantum gravity will still resemble the (thermal) reduced state of the bulk Rehren theory on the right wedge of the maximally extended Schwarzschild-AdS -- see Section 6.4 in \cite{kay2014brick} and Section 7.3 in \cite{kay2015instability}.  (Of course, a similar situation will hold for the left wedge and the left boundary cylinder, but, in full AdS/CFT, that will be entirely unconnected from anything that happens on the right wedge and the right boundary cylinder and therefore they may as well not to be there as far as the right wedge and right boundary cylinder are concerned.)  

The main lesson from all this for AdS/CFT when there is no black hole in the bulk is that one still expects it to be just the matter degrees of freedom in the bulk which are equivalent to the boundary CFT algebra -- in justification of what we wrote in Point (4) above.   The difference presumably is that (as stated in the last sentence of the penultimate paragraph of Section 6.4 of \cite{kay2014brick}) the state of full quantum gravity in the bulk will, in the absence of a black hole, be (approximately) unentangled between matter and gravity so that a pure state on the total matter-gravity theory in the bulk will give rise to a reduced state of the matter alone which is also (approximately) pure.

It seems worth adding that there seem to be a number of clues in the string theory literature which would seem to fit with the above (and indeed to fit with our wider matter-gravity entanglement hypothesis \cite{kay2015entropy}):   First we would cite the paper of Giddings \cite{giddings2013string} which we also mentioned in Section 6.4 of \cite{kay2014brick} which gives independent arguments in favour of the map from bulk to boundary in AdS/CFT being a one-way map.\footnote{We would insert a brief remark here about another paper of Giddings \cite{giddings2015hilbert} that proposes a role for operator algebras and, in particular, for type III$_1$ von Neumann factors in quantum gravity. (Type III$_1$ factors have also been invoked in connection with quantum gravity in recent work \cite{emelyanov2015holography} of Emelyanov.)   Our point of view is that while operator algebras and  type III$_1$ factors clearly do play an important role in quantum field theory in curved spacetime, it is much less obvious to us that they have any role to play in quantum gravity.  In our own work on our matter-gravity entanglement hypothesis (see again \cite{kay2015entropy} for details) we have assumed that we may work with the standard Hilbert space approach to quantum mechanics -- assuming, in particular, that there is a total Hilbert space which arises as a tensor product of a matter Hilbert space and a gravity Hilbert space,  that in some sense there is unitary time-evolution and that we may identify the notion of `state' with `density operator'.}  

Secondly we wish to refer to this passage from a recent paper of McGough and Verlinde:

\medskip

``\textit{this interpretation immediately raises an important
puzzle,} $\dots$.

\textit{According to the usual AdS/CFT dictionary, any typical CFT state with large enough
energy describes a black hole in the bulk. The level density of the CFT indeed matches the
microscopic B-H entropy. However, to write a state with entanglement entropy proportional
to $S_{BH}$, one needs to include two Hilbert space sectors each with entropy at least equal to $S_{BH}$. The CFT seems to provide only one of these sectors. So where is the other sector?}''

\medskip
\noindent
(for the full context and meaning of which, we must, for lack of space, refer the reader to paper \cite{mcgough2013bekenstein} in which it occurs). 
 
On our view, we seem to be in a position to answer this puzzle by replying that the two sought-for Hilbert space sectors are the bulk matter sector and the bulk gravity sector.
(See also the critique of the mainstream understanding of black hole entropy in string theory in \cite{kay2012modern} as also quoted in \cite{kay2015entropy} and see the discussion of what the matter-gravity split might correspond to in string theory in \cite{kay2012modern} and \cite{kay2012more}.)

Finally, we point that that, even though we have argued that, in AdS/CFT, the boundary CFT is only equivalent to the matter sector of the bulk quantum gravity theory, that of course does not mean that purely gravitational features of the bulk will not be reflected in features of the boundary theory.    The reason why not, in very crude terms, is that if one knows everything just about the matter degrees of freedom of the bulk, then, by Einstein's equations, one may infer a lot about the state of the gravitational field too.\footnote{This remark, and some of our other remarks at the end of our Point (6) are relevant to our matter-gravity entanglement hypothesis in general.}

(7) One might ask:  If the boundary CFT is equivalent to a quantum theory (not involving gravity) on the fixed AdS background spacetime, how can it be that the formula for the entanglement entropy of a pair of complementary Rehren wedges involves Newton's constant, $G$ (in the bulk)?   To answer this, we note first if we view our boundary-cylinder as a (d+1-dimensional) spacetime in its own right, it can actually be regarded as the conformal boundary of any one of a one-parameter family of AdS interiors, each with a different radius $R$ (and solving the Einstein equations for the value of the cosmological constant, $\Lambda$, related to $R$ by the BH relation).   There is a separate Rehren bijection for each value of $R$ and, say for $d=1$, for a given boundary CFT and an arbitrary such value of $R$, and a given pair of complementary balls at equal time on our boundary cylinder, the entanglement entropy between the two complementary bulk Rehen wedges which correspond to our balls will (of course) be $1/4G'$ times the area of their shared ridge, where $G'$ is related to the parameters of our CFT by $G'={3R\over 2c}$.   Only for the particular value of $R$ which satisfies the BH relation (\ref{BrownHenneaux}) for the true $G$ will the entanglement entropy have the correct value: $1/4G$ times the area of the shared ridge.  This does not contradict our view that the AdS/CFT correspondence is an equivalence between a subtheory (i.e.\ the matter sector only) of the bulk theory and the boundary CFT.  On the contrary, it is consistent with the  expectation that only for the value of $R$ which satisfies the BH relation, will there be any chance at all that a consistent bulk theory involving both gravity and matter can exist whose matter sector is equivalent to the boundary CFT.   (This replacing the conventional expectation that only for the value of $R$ which satisfies the BH relation, will there be any chance at all that a consistent bulk theory involving both gravity and matter can exist which is fully equivalent to the boundary CFT.)  A possibly helpful slogan to summarize this paragraph is:   \textit{It is not Rehren's algebraic holography which knows about the  value of Newton's constant; rather, it is the BH relation that knows about the value of Newton's constant.}  (Similarly, as we saw in \cite{kay2014brick} [see the last paragraph in Section 6.2 there] when one considers holography involving a thermal state on the CFT boundary and a black hole in the bulk, then the `fixed background holography' considered there equally doesn't know about the mass of the black hole; instead it is the appropriate AdS/CFT relation between the parameters of the CFT [now including the temperature of the thermal state on the boundary] and the mass and asymptotic AdS radius which determines the correct possible values of these. )

In conclusion, we have argued that a comparison between Rehren's algebraic holography theorem and Ryu-Takayanagi's work on entanglement entropy naturally leads to a strengthening of our previous arguments that, in AdS/CFT, the boundary CFT is equivalent, not to the full bulk quantum gravity theory, but just to its matter sector.  Moreover, we have given further evidence for the validity of the Bianchi-Myers conjecture -- quoted as the italicized statement in Point (5) above -- but with the important replacement of the phrase ``the degrees of freedom'' by ``the \textit{matter} degrees of freedom''.

It would obviously be of interest to have concrete examples of Rehren duals of boundary CFTs which occur in standard AdS/CFT correspondences.  (The only existing concrete example of Rehren's algebraic holography of which I am aware [see \cite{rehren2000algebraic} and \cite{kay2008pre}] has, as the bulk field, the covariant Klein-Gordon equation.   The CFT to which it is dual [it is a certain massless generalized free field] is, however, pathological both in that it fails to satisfy the time-slice condition and also has anomalous thermodynamic properties.)  According to our arguments here and in \cite{kay2014brick}, such a Rehren bulk theory will give an approximate description of the matter sector in full AdS/CFT.

\end{document}